\documentclass[%
reprint,
superscriptaddress,
 amsmath,amssymb,
 aps,
prl,
floatfix,
]{revtex4-1}

\usepackage{graphicx}
\usepackage{hyperref}
\usepackage[usenames,dvipsnames]{xcolor}

\DeclareMathOperator{\Tr}{Tr}

\definecolor{blue(pigment)}{rgb}{0.2, 0.2, 0.6}
\definecolor{darkerblue}{rgb}{0.0, 0.0, 0.4}
\definecolor{darkblue}{rgb}{0.0,0.0,0.5}
\definecolor{darkgreen}{rgb}{0.0,0.4,0.0}
\hypersetup{
   colorlinks,
   linkcolor=blue(pigment),
    citecolor=darkgreen,
   urlcolor=darkblue
}

\begin{document}

\title{Non-local emergent hydrodynamics in a long-range quantum spin system}%
\author{Alexander Schuckert}\email{alexander.schuckert@tum.de}
\affiliation{Department of Physics and Institute for Advanced Study, Technical University of Munich, 85748 Garching, Germany}
\affiliation{Munich Center for Quantum Science and Technology (MCQST), Schellingstr. 4, D-80799 M\"unchen}
\author{Izabella Lovas}
\email{izabella.lovas@tum.de}
\affiliation{Department of Physics and Institute for Advanced Study, Technical University of Munich, 85748 Garching, Germany}
\affiliation{Munich Center for Quantum Science and Technology (MCQST), Schellingstr. 4, D-80799 M\"unchen}
\author{Michael Knap}
\email{michael.knap@ph.tum.de}
\affiliation{Department of Physics and Institute for Advanced Study, Technical University of Munich, 85748 Garching, Germany}
\affiliation{Munich Center for Quantum Science and Technology (MCQST), Schellingstr. 4, D-80799 M\"unchen}
\date{\today}

\begin{abstract}
Generic short-range interacting quantum systems with a conserved
quantity exhibit universal diffusive transport at late times. We employ non-equilibrium quantum field theory and semi-classical phase-space simulations to show how this universality is replaced by a more general transport  process in a long-range XY spin chain at infinite temperature with couplings decaying algebraically with
distance as $r^{-\alpha}$. While diffusion is recovered for $\alpha>1.5$, longer-ranged couplings with $0.5<\alpha\leq1.5$
give rise to effective classical L\'evy flights; a random walk with step sizes 
drawn from a distribution with algebraic tails. We find
that the space-time dependent spin density profiles are self-similar,
with scaling functions given by the stable symmetric distributions. As a
consequence, for $0.5<\alpha\leq1.5$ autocorrelations show hydrodynamic tails decaying in time as $t^{-1/(2\alpha-1)}$ and linear-response theory breaks down.  Our findings can be readily verified with current trapped ion experiments.

\end{abstract}

\maketitle
In quantum many-body systems, macroscopic inhomogeneities in a conserved quantity must be transported across the whole system to reach an equilibrium state. As this is in general a slow process compared to local dephasing,  essentially \emph{classical} hydrodynamics is expected to emerge at late times in the absence of long-lived quasi-particle excitations~\cite{PhysRevB.73.035113, lux_hydrodynamic_2014, bohrdt_scrambling_2017,  medenjak_diffusion_2017,1702.08894,PhysRevX.8.031057,PhysRevX.8.031058,parker_universal_2018}.
The universality of this effective classical description may be understood from the central limit theorem: in the regime of incoherent transport, short range interactions lead to an effective random walk with a finite variance of step sizes, leading to a Gaussian distribution at late times. This universality is only broken when quantum coherence is retained, such as in integrable models~\cite{PhysRevX.6.041065,PhysRevLett.117.207201,gopalakrishnan_kinetic_2019,PhysRevLett.122.150605,PhysRevLett.122.210602,PhysRevB.98.125119,PhysRevB.99.121110} or in the vicinity of a many-body localized phase, where rare region effects give rise to subdiffusive transport~\cite{agarwal_anomalous_2015,PhysRevLett.114.100601, PhysRevX.5.031032,PhysRevX.5.031033,bordia_probing_2017,gopalakrishnan_noise-induced_2017,agarwal_rare-region_2017}.

\begin{figure}[ht!]
	\includegraphics[width=\columnwidth]{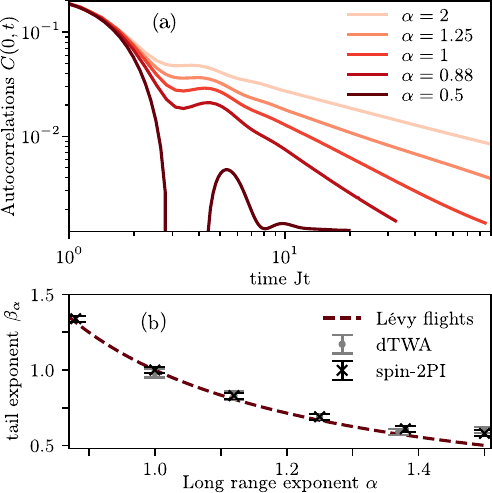}
	\caption{\textbf{Hydrodynamic tails in the spin autocorrelator.} (a) For long range coupling exponents $\alpha>0.5$,  autocorrelations decay algebraically at late times with an exponent that depends on $\alpha$. By contrast for $\alpha\leq 0.5$ hydrodynamic tails are absent.
	(b) The exponents $\beta_\alpha$ of the hydrodynamic tail obtained  from two different approaches (symbols) agree with the predictions from classical L\'evy flights in the thermodynamic limit (dashed curve). Deviations at large $\alpha$ are due to finite time corrections to scaling which can also be understood from L\'evy flights.
	\label{fig_hydrotails}}
\end{figure}

In this work, we show how this universal diffusive transport in short range interacting systems is replaced by a more general, non-local effective hydrodynamical description in systems with algebraically decaying long-range interactions. We use semi-analytical non-equilibrium quantum field theory calculations (referred to as spin-2PI below) and a discrete truncated Wigner  approximation (dTWA)  to show that in a long range interacting XY spin chain, spin transport at infinite temperature effectively obeys a classical master equation with \emph{long-range}, algebraically decaying transition amplitudes. This effective description can be reformulated as a classical random walk with infinite variance of step sizes, giving rise to a generalized central limit theorem and to a late-time description in terms of classical  L\'evy flights~\cite{zaburdaev_levy_2015}, an example for \emph{super}diffusive anomalous transport. As a result, we demonstrate that the full spatio-temporal shape of the correlation function $C(j,t) = \langle \hat{S}_j^z(t)\hat{S}_0^z(0)\rangle$, and in particular, the exponent of the hydrodynamic tail in the autocorrelation function $C(j=0,t)$, depends strongly on the long-range exponent $\alpha$. While for $\alpha>1.5$ we recover classical diffusion, the autocorrelation function shows hydrodynamic tails with an exponent $1/(2\alpha-1)$ for $0.5<\alpha\leq1.5$ as we show in Fig.~\ref{fig_hydrotails}. Furthermore, $C(j,t)$ possesses a self-similar behavior, with the scaling function covering all stable symmetric distributions as a function of $\alpha$, smoothly crossing over from a Gaussian at $\alpha=1.5$ over a Lorentzian at $\alpha=1$ to an even more sharply peaked function as $\alpha\rightarrow 0.5$. We also extract the generalized diffusion coefficient $D_\alpha$ from the scaling functions, and explain its $\alpha$ dependence by L\'evy flights; quantum effects are incorporated in a many-body time scale depending only weakly on $\alpha$. For $\alpha\leq0.5$ no emergent hydrodynamic behavior is found as the system relaxes instantaneously in the thermodynamic limit~\cite{PhysRevLett.110.170603}.

This work not only shows how non-local transport phenomena emerge in long-range interacting systems, but also establishes both nonequilibrium quantum field theory and discrete truncated Wigner simulations as efficient tools to study transport phenomena in the thermalization dynamics of quantum many-body systems.

\textbf{Model.--}We study the long range interacting quantum XY chain  with open boundary conditions, given by the Hamiltonian
\begin{equation}
\hat H = -\frac{1}{2}\sum_{i\neq j=-L/2}^{L/2}\frac{J}{\mathcal{N}_{L,\alpha}|i-j|^\alpha}\left(\hat{S}^x_i\hat{S}^x_j+\hat{S}^y_i\hat{S}^y_j\right).
\label{eq:XY_Ham}
\end{equation}
Here, $\hat{S}^\alpha=\frac{1}{2}\hat\sigma^\alpha$ denotes spin-$\frac{1}{2}$ operators given in terms of Pauli matrices, $L$ is the (odd) length of the chain~\footnote{We always assume integer divisions when we write $\frac{L}{2}$.}, and we set $\hbar=1$. The interaction strength $J$ is rescaled with the factor $\mathcal{N}_{L,\alpha}=\sqrt{\sum_{j\neq 0} \left|j\right|^{-2\alpha}}$ in order to remove the $L$ and $\alpha$ dependence of the time scale associated with the perturbative short time dynamics of the central spin at $i=0$. The above model shows chaotic (Wigner-Dyson) level statistics for the whole range of $\alpha$ considered here ($0.5\leq\alpha\leq2$) and is an effective description of the long range transverse field Ising model for large fields~\cite{jurcevic_quasiparticle_2014, smith_many-body_2016}. In particular, it conserves the total $S^z$ magnetization, with product states in the $S^z$ basis evolving radically differently depending on the complexity of the corresponding magnetization sector. For just a few spin flips on top of the completely polarized state, the dynamics can be exactly solved and are described in terms of ballistically propagating spin waves, with a diverging group velocity at $\alpha=1$~\cite{Hauke2013,Richerme2014,jurcevic_quasiparticle_2014} related to the algebraic leakage of the Lieb-Robinson bound~\cite{Lieb1972,Hastings2006,Eisert2013,Tran2018}.  In contrast,  here we show that the exponentially large Hilbert space sector for an extensive number of spin flips gives rise to rich transport phenomena, driven by the long-range nature of the interactions.

\textbf{Effective stochastic description of long-range transport.--}  As the model in Eq.~\ref{eq:XY_Ham} is equivalent to long-range hopping hard core bosons, we conjecture the effective classical equation of motion for the transported local density $f_j(t)$, in our case $\left\langle \hat S^z_j(t)\right\rangle+\frac{1}{2}$, to be of the form \footnote{A Master equation in terms of a local \emph{probability} density may be obtained by normalizing the local density.}
\begin{align}\label{eq:master}
\partial_t f_j(t) &= \sum_{i\neq j} \left( W_{i\rightarrow j}\, f_i (1-f_j) - W_{j\rightarrow i}\, f_j (1-f_i)\right).
\end{align}
Here, the transition rate $W_{i\rightarrow j}$ is determined by the microscopic transport processes present in the Hamiltonian, in our case the long-range hopping of spins. More specifically, from Fermi's golden rule the transition rate for a flip flop process between spins $i$ and $j$  is proportional to $|\langle \uparrow_i\downarrow_j|\hat H|\downarrow_i\uparrow_j\rangle|^2$, and hence we phenomenologically set
\begin{equation}
W_{i\rightarrow j}=W_{j\rightarrow i}=\frac{\lambda}{|i-j|^{2\alpha}},
\end{equation}
where $\lambda^{-1}$ is a characteristic time scale determined by the full many-body Hamiltonian.

Starting from an initial state with a single excitation in the center of the chain, the solution of this Master equation is given by~\cite{supplement}
\begin{equation}
f_j(t)\approx\Bigg\{\begin{array}{lr}
(D_\alpha t)^{-1/2}\,G\left(\frac{|j|}{(D_\alpha t)^{1/2}}\right)        & \text{for } \alpha>1.5\\
(D_\alpha t)^{-\beta_\alpha}F_\alpha\left(\frac{|j|}{(D_\alpha t)^{\beta_\alpha}}\right)        & \text{for } 0.5<\alpha\leq 1.5
        \end{array}
\label{eq:scaling_form}
\end{equation}
in the limit of long times and large system sizes. Here, $G(y)=\exp(-y^2/4)/8\sqrt{\pi}$ denotes the Gaussian distribution,  indicating normal diffusion for $\alpha>1.5$ with diffusion constant $D_\alpha\propto \lambda$. For $0.5<\alpha\leq 1.5$, $G(y)$ is replaced by the family of stable, symmetric distributions $F_\alpha(y)$, given by
\begin{equation}\label{eq:Fy}
F_\alpha(y)=\dfrac{1}{4}\int\dfrac{\mathrm{d}k}{2\pi}\exp(-|k|^{1/\beta_\alpha})\exp(iyk),
\end{equation} 
with the constant prefactor $D_\alpha=\lambda c_\alpha$ constituting a generalized diffusion coefficient \footnote{The prefactor 1/4 accounts for the normalization of the correlation function, $C(j,t=0)=\delta_{0,j}/4$. Furthermore, reinstating a lattice spacing $a$, the units of $D_\alpha$ depend on $\alpha$, in particular it is a velocity for $\alpha=1$.}. We find $c_\alpha = -2\Gamma(1-2\alpha)\sin(\pi\alpha)$ from the classical Master equation, with $\Gamma$ denoting the gamma function~\cite{supplement}.  
Furthermore, the exponent of the hydrodynamic tail $\beta_\alpha$ is given by
\begin{equation}
\beta_\alpha = \dfrac{1}{2\alpha-1}.
\end{equation}
 The Fourier transform in Eq.~\eqref{eq:Fy} only leads to elementary functions for $\alpha=3/2$ and $\alpha=1$, resulting in a Gaussian and a Lorentzian distribution, respectively~\footnote{Other closed form solutions exist~\cite{Lee}, for example for $\alpha=1.25$ in terms of hypergeometric functions.}. The scaling functions $F_\alpha(y)$ are the fixed point distributions in the generalized central limit theorem~\cite{Gnedenko1954} of i.i.d. random variables with heavy tailed distributions. Importantly, $F_\alpha(y)$ has  diverging variance for $\alpha<1.5$, undefined mean for $\alpha\leq 1$, and displays heavy tails $\sim |y|^{-2\alpha}$. The classical Master equation hence predicts a cross-over from diffusive ($\alpha\geq1.5$) over ballistic ($\alpha=1$) to super-ballistic ($0.5<\alpha<1$) transport.

When adding a linear magnetic field gradient $\sim E\sum_i i \hat S^z_i$ to the Hamiltonian, the resulting classical Master equation predicts the spin current to depend non-linearly on the arbitrarily weak $E$ for $\alpha<1.5$, indicating a breakdown of linear response theory~\cite{arkhincheev_nonlinear_2001,supplement}. Calculating the current response function from Eq.~\ref{eq:scaling_form}, we find a diverging response for vanishing momentum $q\rightarrow 0$ for every value of the frequency $\omega$~\cite{supplement,Forster}.   

\textbf{Quantum dynamics from spin-2PI and dTWA.--}In the following, we demonstrate the emergence of these effective classical dynamics in the quantum dynamics of the Hamiltonian~\eqref{eq:XY_Ham}, by studying the unequal time correlation function
\begin{equation}
C(j,t):=\Tr\left[\hat S^z_{j}(t)\hat S^z_{0} (0)\right]_{\mathrm{|j=0\rangle=|\uparrow\rangle}}.
\label{eq:def_corr_fct}
\end{equation}
Here, we perform the trace over product states in the $S^z$ basis, restricted to the Hilbert space sector with $\sum_i  S_i^z=\frac{1}{2}$, such that $\langle S^z_i(t=0)\rangle=\frac{1}{2}\,\delta_{0,i}$ for all spins $i$. 
This way, we probe the transport of a single spin excitation moving in an infinite temperature bath with vanishing total magnetization. 

We employ two complementary, approximate methods to study the dynamics at long times and for large system sizes, in a regime that is challenging to access by numerically exact methods~\cite{PhysRevA.99.032114}. Schwinger boson spin-2PI~\cite{PhysRevX.5.041005, PhysRevB.98.224304}, a non-equilibrium quantum field theory method, employs an expansion in the inverse coordination number $1/z$ to reduce the many-body problem to solving an integro-differential equation that scales algebraically in system size. As the effective coordination number is large in a long-range interacting system, we expect this approximation to be valid for small $\alpha$. The discrete truncated Wigner approximation evolves the classical equations of motion, while introducing quantum fluctuations by sampling initial states from the Wigner distribution~\cite{wootters_wigner-function_1987,polkovnikov_phase_2010,rey_2015,kastner_2016} and was shown to be particularly well suited for studying long-range interacting systems~\cite{rey_2015,PhysRevLett.120.063601}. In both methods, we evaluate $C(j,t)$ by starting from random initial product states in the $S^z$ basis and then averaging over sufficiently many such initial states~\footnote{In spin-2PI simulations we used $4-16$ different initial states, while in dTWA the averaging over initial states is performed in parallel with the Monte Carlo averaging over the Wigner distribution; here we typically use $\sim 10^5$ samples.}. If not stated otherwise, all our results have been converged with respect to system size, for which we employed chains with $201-601$ sites.

We study two distinct regimes in the dynamics. A perturbative short time regime characterized by initial dephasing is followed by the emergent effective classical long-range transport described by the Master equation. 

\begin{figure}[t!]
\includegraphics{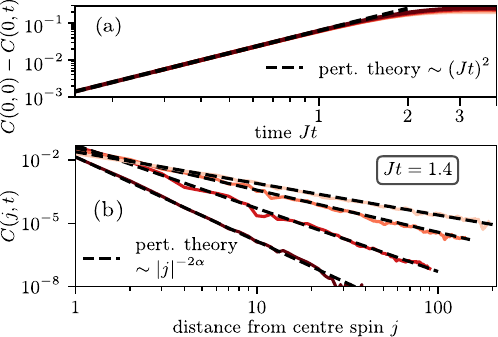}
\caption{\textbf{Short time dynamics.} We compare spin-2PI results with second order perturbation theory, Eq.~\ref{eq:perttheory}. (a) The collapse of the autocorrelator for different exponents $\alpha$ shows that the short time evolution is independent of $\alpha$ and $L$ when the Hamiltonian is rescaled with $\mathcal{N}_{\alpha,L}$. (b) The un-equal-time correlation function for $\alpha\in\lbrace0.75,1,1.5,2\rbrace$ (from top to bottom), shows algebraic tails that are entirely captured by second order perturbation theory. We used a moving average over $5-10$ lattice sites to smoothen the results. \label{fig:shorttimes}}
\end{figure}

\textbf{Perturbative short time dynamics.--} At short times, second-order perturbation theory yields
\begin{equation}
\Tr\left[ \hat S^z_{j}(t)\hat S^z_{0} (0)\right] \approx \bigg\{\begin{array}{lr}
        \frac{1}{4}(1-\frac{J^2t^2}{4})& \text{for } j=0\\
        \left(\frac{J t}{4\mathcal{N}_{L,\alpha}}\right)^2\frac{1}{|j|^{2\alpha}} & \text{for } j\neq 0
        \end{array}.
\label{eq:perttheory}
\end{equation}
Physically, in this regime each spin is precessing in the effective magnetic field created by all other spins. The autocorrelation function is independent of $\alpha$ and $L$ due to our choice of the  normalization factor $\mathcal{N}_{L,\alpha}$, ensuring that the typical magnetic field at the center of the chain remains of the order of $J$.
The spatial correlation function at a fixed time inherits the algebraic behavior of the interaction strength, falling off as $|j|^{-2\alpha}$ between spins of distance $j$, which is reproduced by both dTWA (not shown) and spin-2PI, see Fig.~\ref{fig:shorttimes}.

\begin{figure*}[t!]
\includegraphics{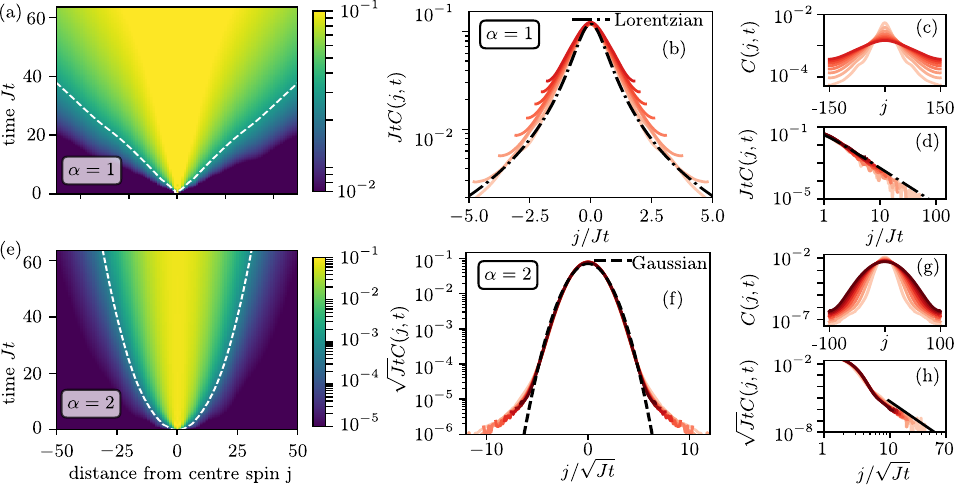}
\caption{\textbf{Emergent self-similar time evolution.} The correlation function $C(j,t)$ obtained from spin-2PI for chains of lengths $L=201$ ($\alpha=2$, subfigures (e-h)), $L=301$ ($\alpha=1$, subfigures (a-d)). (a,e) $C(j,t)$ multiplied by $t^{1/(2\alpha-1)}$ to account for the overall decay expected from L\'evy flights shows a diffusive cone for $\alpha=2$, whereas for $\alpha=1$ a ballistic light-cone emerges. The contour lines for $\alpha=1,2$ correspond to values $t^{1/(2\alpha-1)}C(j,t)=0.03,10^{-4}$, respectively. (b,f) Rescaling of linearly spaced time slices for $23\leq Jt\leq 84$ $(\alpha=1)$ and $42\leq Jt\leq 226$ $(\alpha=2)$ (lines become darker as time increases) for the same data as in (a,e) agrees well with the scaling function expected from classical L\'evy flights, Eq.~\ref{eq:scaling_form}. The only fitting parameter is the generalized diffusion coefficient. (d,h) Rescaled time slices ($2\leq Jt\leq 28$) on a double-logarithmic scale reveal for $\alpha=1$ the heavy tail $\sim y^{-2}$ expected from L\'evy flights (Eq.~\ref{eq:scaling_form}), where the dashed-dotted line is the same fit as in (b). The tail $\sim y^{-4}$ (thick black line) for $\alpha=2$ ($8\leq Jt\leq 85$) is a finite time effect also present in classical L\'evy flights. (c,g) Unscaled data. \label{fig_colorplot_scaling}}
\end{figure*}

\textbf{Hydrodynamic tails.--} The scaling form from classical L\'evy flights in Eq.~\ref{eq:scaling_form} implies the presence of a hydrodynamic tail in the autocorrelation function $C(j=0,t)$ with exponent $\beta_\alpha = 1/(2\alpha-1)$, which replaces the universal exponent $1/2$ for diffusion in 1D, see Fig.~\ref{fig_hydrotails} for our field theory results. For $\alpha\rightarrow 1.5$ we find slight deviations from $\beta_\alpha$, these can however be explained by a subtle finite-time effect also present in classical L\'evy flights~\cite{supplement}. For $\alpha<0.5$ we find no hydrodynamic tail for the numerically accessible system sizes $L<601$. This matches the expectation that the system relaxes instantaneously in the thermodynamic limit~\cite{PhysRevLett.110.170603}, which is also indicated by the fact that the perturbative short time scale diverges, $\mathcal{N}_{L\rightarrow \infty,\alpha}= \infty $, for $\alpha\leq0.5$. On even longer time scales, the discretized Fourier transform underlying the derivation of Eq.~\ref{eq:scaling_form} is dominated by the smallest wavenumber in finite chains, and the hydrodynamic tail is replaced by an exponential convergence towards  the equilibrium value $0.25/L$ with a rate $\sim (1/L)^{2\alpha-1}$.

\textbf{Self-similar time evolution of correlations.--} In Fig.~\ref{fig_colorplot_scaling} we show the spreading of $t^{\beta_\alpha}C(j,t)$ for two values of $\alpha$. While for $\alpha=2$ a diffusive cone is visible, the spreading for $\alpha=1$ is ballistic as expected from the Master equation. The scaling collapse of these data shows good agreement with classical L\'evy flights, Eq.~\ref{eq:scaling_form}, at late times. Interestingly, we find heavy tails even for $\alpha\geq1.5$. We explain these by sub-leading corrections to the scaling ansatz Eq.~\ref{eq:scaling_form} present in the Master equation~\cite{supplement}. They survive up to algebraically long times for $\alpha>1.5$,  turning to a logarithmic correction at $\alpha=1.5$ ~\cite{zarfaty_infinite_2019}.

For $\alpha\gtrsim 2$ we furthermore find signs of peaks propagating ballistically for intermediate times in the dTWA scaling functions, which survive longer as $\alpha$ increases. These peaks are remnants of the integrable point at $\alpha=\infty$~\cite{supplement}. Such behaviour is not present in the spin-2PI data as this method is not able to capture integrable behaviour~\cite{PhysRevB.98.224304}.
 
\begin{figure}[b!]
\includegraphics[width=\columnwidth]{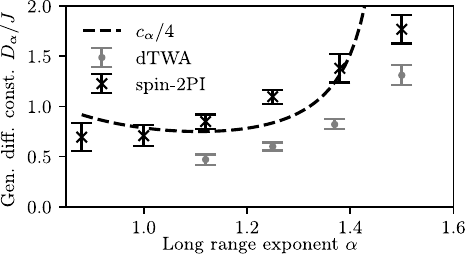}
\caption{\textbf{Generalized diffusion constant.} The $\alpha$ dependence of the diffusion constant obtained from fits with the  scaling function of  L\'evy flights, Eq.~\ref{eq:scaling_form}. The qualitative behavior follows the L\'evy flight prediction $D_\alpha\sim c_\alpha$ for $\alpha<1.5$. \label{fig:diffconst}}
\end{figure}

\textbf{Generalized diffusion constant.--} The only free parameter of our effective classical description is the generalized diffusion coefficient $D_\alpha$, which we obtain from the fits to the scaling function. In Fig.~\ref{fig:diffconst} we  show that the leading  $\alpha$ dependence of $D_\alpha$ can be explained by $D_\alpha\sim c_\alpha$ for $\alpha<1.5$~\footnote{The spurious divergence of $c_\alpha$ in the limit $\alpha=1.5$ is cured by logarithmic corrections to scaling~\cite{supplement}. Furthermore, without the normalisation $N_{L,\alpha}$,  $D(\alpha)\sim N_{L,\alpha}\rightarrow\infty$ for $\alpha\rightarrow 0.5$.}, hence  the prefactor $\lambda^{-1}$, constituting the \emph{quantum} many-body time scale,  depends only weakly on $\alpha$.  As expected from their differing approximate treatment of the quantum fluctuations in the system, we find slight differences between the values of $\lambda$  determined by spin-2PI and dTWA, $\lambda_\mathrm{2PI}\approx 0.25$ and $\lambda_\mathrm{dTWA}\approx 0.15$. For $\alpha > 1.5$ we find considerable differences between the dTWA and spin-2PI results, because the emergent  ballistic peaks, stemming  from the nearby integrable point, accelerate the spreading in the dTWA simulations.

\textbf{Conclusions.--} In this paper, we have shown that spin transport at high temperatures in long-range interacting XY-chains is well described by L\'evy flights for long-range interaction exponents $0.5<\alpha\leq1.5$, effectively realizing a random walk with infinite variance of step sizes. In particular, we have shown that the scaling function of the unequal time spin correlation function covers the stable symmetric distributions, in accordance with the generalized central limit theorem. While the system relaxes instantly for $\alpha<0.5$, standard diffusion was recovered for $\alpha>1.5$, with heavy tails from finite time corrections surviving until extremely long times. We demonstrated the non-trivial dependence of the generalized diffusion coefficient $D_\alpha$ on $\alpha$, and found that it is captured by classical L\'evy flights, with the quantum many body time scale being approximately independent of $\alpha$. While we only studied one-dimensional systems, we expect this phenomenon to generalize straightforwardly to $d>1$ dimensions. Assuming the effective classical L\'evy flight picture persists, superdiffusive behaviour would be found for $d/2<\alpha<1+d/2$ with the exponent of the hydrodynamic tails given by $d/(2\alpha-d)$~\cite{supplement}. Furthermore, we indicated that L\'evy flights also imply a non-linear response of the spin current to magnetic field gradients. 

The long-range transport process found here can be experimentally studied in current trapped ion experiments~\cite{PhysRevLett.92.207901}, which can reach the required time scales~\cite{brydges_probing_2018, zhang_observation_2017, PhysRevX.8.021012,PhysRevLett.122.050501}. The effective infinite temperature states can also be realized by sampling over random product states which are then evolved in time.

\begin{acknowledgments}
\textbf{Acknowledgments.--}We thank Rainer Blatt, Eleanor Crane, Iliya Esin, Philipp Hauke, Christine Maier, Asier Pi\~neiro Orioli, Tibor Rakovszky and Achim Rosch for insightful discussions and the Nanosystems Initiative Munich (NIM) funded by the German Excellence Initiative for access to their computational resources. We acknowledge support from the Max Planck Gesellschaft (MPG) through the International Max Planck Research School for Quantum Science and Technology (IMPRS-QST), the Technical University of Munich - Institute for Advanced Study, funded by the German Excellence Initiative, the European Union FP7 under grant agreement 291763, the Deutsche Forschungsgemeinschaft (DFG, German Research Foundation) under Germany's Excellence Strategy--EXC-2111--390814868, the European Research Council (ERC) under the European Union's Horizon 2020 research and innovation programme (grant agreements
No 771537 and 851161), from DFG grant No. KN1254/1-1, and DFG TRR80 (Project F8).
\end{acknowledgments}

\begin{widetext}
\section{SUPPLEMENTARY MATERIAL}

\onecolumngrid 
\section{Scaling functions from the classical master equation}

 Here we derive the scaling collapse presented in the main text from the master equation
\begin{equation}
\partial_t f_j = \sum_{i\neq j} W_{ij}(f_i-f_j),
\label{eq:ME}
\end{equation}
with transition rates
\begin{equation} 
W_{ij}=\dfrac{\lambda}{|i-j|^{2\alpha}}.
\end{equation}
By taking the Fourier transform of this equation, we arrive at
\begin{equation}
\partial_t f(k) = [W(k)-W(k=0)]\, f(k),
\end{equation}
where
\begin{equation}
f(k) = \sum_{j=-L/2}^{L/2} e^{-ikj}f_j,
\end{equation}
and
\begin{align}
W(k)-W(k=0)&=\lambda\left[\sum_{j=-L/2}^{-1}+\sum_{j=1}^{L/2}\right]\left(e^{-ikj} - 1\right) / |j|^{2\alpha}\approx2\, \lambda\int_1^{L/2} dx \,\left(\cos{kx} - 1\right) / x^{2\alpha},
\label{eq:W_K}
\end{align}
with $k=2\pi n/L$, $n=-L/2,...,L/2$. 

The long time behavior is dominated by large wavelengths $k\ll 1$. For $0.5<\alpha<1.5$ the integral remains convergent when we remove both the upper and lower cutoffs, leading to the following approximation in the regime $k\ll 1$,
\begin{align}\label{eq:rates}
W(k)-W(k=0)&\approx 2\,\lambda\int_0^{L/2} dx \,\left(\cos{kx} - 1\right) / x^{2\alpha}-2\,\lambda\int_0^{1} dx \,\left(\cos{kx} - 1\right) / x^{2\alpha}\nonumber\\ 
&\approx 2\,\lambda\int_0^{\infty} dx \,\left(\cos{kx} - 1\right) / x^{2\alpha} + \lambda\,k^2 \,\int_0^{1} dx\, x^{2-2\alpha}\nonumber\\ 
&=\left(-c_\alpha\,|k|^{2\alpha-1} + \dfrac{k^2}{3-2\alpha}\right)\lambda.
\end{align}
Here 
$$c_\alpha=-2\int_0^{\infty} dz \,\left(\cos{z} - 1\right) / z^{2\alpha}=-2\,\Gamma(1-2\alpha)\sin(\alpha\pi),$$
with $\Gamma$ denoting the gamma function. Note that $c_\alpha>0$ for $0.5<\alpha<1.5$.

For $0.5<\alpha<1.5$, the first term in Eq.~\eqref{eq:rates}, $\sim |k|^{2\alpha-1}$ , will dominate the long time behavior, leading to
\begin{equation}
\partial_t f(k) \approx -\lambda c_\alpha\,|k|^{2\alpha-1}\quad\Rightarrow\quad f(k,t)=f(k,0)\,e^{-\lambda c_\alpha\,|k|^{2\alpha-1} t}.
\end{equation} 
In particular, taking an initial state with a single localized excitation, $f_j(t=0)=\delta_{j,0}/4$ and hence $f(k,0)\equiv 1/4$ with the factor $1/4$ stemming from $(\hat S^z)^2=1/4$, we arrive at the scaling ansatz 
\begin{equation}
f_j(t)\approx \dfrac{1}{4}\int\dfrac{dk}{2\pi}\; \exp\left(ikj-\lambda\, t\, c_\alpha\,|k|^{2\alpha-1}\right)=(\lambda c_\alpha t)^{-1/(2\alpha-1)}\,F_\alpha\left(\dfrac{|j|}{(\lambda c_\alpha t)^{1/(2\alpha-1)}}\right),
\label{eq:scalingfinal}
\end{equation}
with $F_\alpha(y)$ given in the main text.

\textbf{Diffusion for $\alpha>1.5$.--} While we used $\alpha<1.5$ in the derivation of Eq.~\ref{eq:rates}, the expression is in fact valid for all $\alpha>0.5$. This can be shown by  evaluating the following integral exactly,
\begin{align*}
W(k)-W(k=0)&\approx2\, \lambda\, |k|^{2\alpha-1}\int_{|k|}^{\infty} dz \,\left(\cos{z} - 1\right) / z^{2\alpha},
\end{align*}
 and expanding the resulting expression around $k=0$. Noting that for $\alpha>1.5$ the $|k|^{2\alpha-1}$ term is subdominant, we arrive at
\begin{equation}
f^{\alpha>1.5}_j(t)\approx \dfrac{1}{4}\int\dfrac{dk}{2\pi}\; \exp\left(ikj-\dfrac{\lambda}{2\alpha-3} k^2 t\right)=\left(D_\alpha  t\right)^{-1/2}\,G\left(\dfrac{|j|}{\left( D_\alpha t\right)^{1/2}}\right),
\label{eq:scalingdiff}
\end{equation}
reproducing diffusive behaviour for $\alpha>1.5$ with diffusion coefficient $D_\alpha=\lambda/(2\alpha-3)$ and a Gaussian $G(y)=\exp(-y^2/4)/8\sqrt{\pi}$. We note that in contrast to the regime $\alpha<1.5$, here the dTWA results for the quantum many-body time scale $\lambda$ show a strong $\alpha$ dependence, with $D_\alpha$ increasing as a function of $\alpha$ due to the approach to the integrable point at $\alpha\rightarrow\infty$.

\textbf{Exponential late time decay of the autocorrelation function.--} For finite system sizes $L$, approximating the discrete Fourier sums by integrals eventually breaks down at very long times. In this regime the time evolution will be dominated by the two smallest non-zero wave-numbers, $k=\pm 2\pi/L$, leading to an exponential decay
\begin{equation}
f_j(t)\approx\dfrac{1}{L}\left[f(k=0,t) + \sum_{k_0 = \pm 2\pi/L}e^{ik_0 j}f(k_0,t)\right]= \dfrac{1}{4L}\left[1 + 2 \cos(2\pi j/L) e^{-\lambda t c_\alpha (2\pi/L)^{2\alpha-1}}\right] 
\end{equation}
for the case of $\alpha\leq1.5$. The exponent of this exponential decay is hence expected to scale with the system size as $\sim (1/L)^{2\alpha-1}$. This prediction is in agreement with our spin-2PI numerical results. Moreover, this result can be used to extract the diffusion coefficient $D_\alpha=\lambda c_\alpha$ from finite size data. 

\begin{figure*}[b!]
\includegraphics{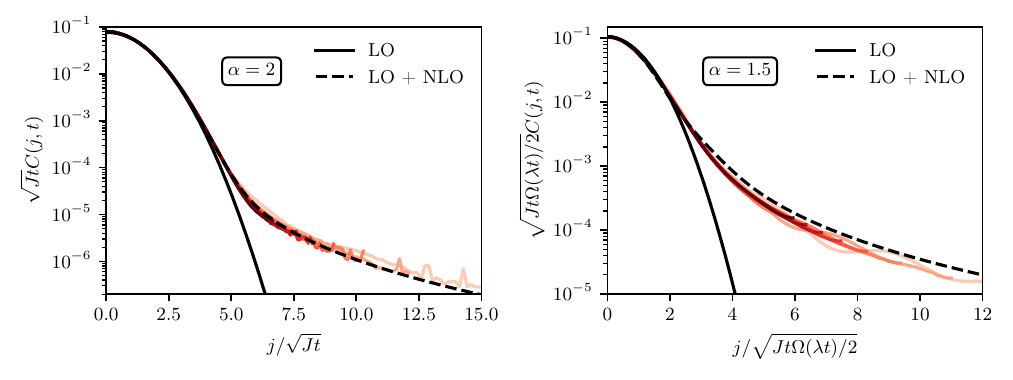}
\caption{\textbf{Corrections to scaling.} The heavy tails found in the scaling functions for $\alpha\geq1.5$ are completely captured by the finite time corrections to scaling in the classical Master equation, with fit functions given in Eqs.~\ref{eq:corr2_v2} and \ref{eq:corr15}. Note that there is only a single fit parameter (given by the quantum many body time scale $1/\lambda$), with its numerical value being approximately equal in the fit to the (scaling) functions obtained from the leading order (LO) and next-to-leading order (LO+NLO) in a simultaneous $k\rightarrow 0$, $t\rightarrow \infty$ expansion.\label{fig:scalingcorr}}
\end{figure*}

\section{Corrections to scaling}
 For finite times, the two terms in Eq.~\ref{eq:rates} compete, leading to corrections to the leading-order scaling ansatz shown in the main text. As we show below, these corrections survive until algebraically long times  for $\alpha>1.5$,  and they add a logarithmic correction to the scaling ansatz at the threshold value $\alpha=1.5$, explaining all major deviations from leading-order scaling  we observe in our numerical data.
 
 \textbf{Diffusive regime, $\alpha>1.5$.--} Taking into account both terms in Eq.~\ref{eq:rates}, then rescaling  as $k\rightarrow k\sqrt{D_\alpha t}$ and $j\rightarrow y=j/\sqrt{D_\alpha t}$ where $D_\alpha = \lambda/(2\alpha-3)$,  and finally expanding the remaining time dependent term for $t\rightarrow\infty$, we arrive at
\begin{align}
\sqrt{D_\alpha t}\,f_j(t)&\approx\dfrac{1}{4}\int\dfrac{dk}{2\pi}\; \exp\left(iky-k^2\right)\left(1-(D_\alpha t)^{-\alpha+3/2}\,c_\alpha (2\alpha-3) |k|^{2\alpha-1} \right)\nonumber\\
&= G(y)-(D_\alpha t)^{-\alpha+3/2}\;\dfrac{c_\alpha (2\alpha-3)\Gamma(\alpha)}{16\,\pi}\,{}_1F_1\left[\alpha,\frac{1}{2},-\frac{y^2}{4}\right],
\label{eq:corr2}
\end{align}
with  ${}_1F_1\left[\cdot,\cdot,\cdot\right]$ denoting the Kummer confluent hypergeometric function. Most importantly, the latter exhibits heavy tails $\sim y^{-2\alpha}$ for $y\rightarrow\infty$, reproducing the finite time data for $\alpha=2$ in Fig.3 of the main text. 

We show this behaviour more explicitly in Fig.~\ref{fig:scalingcorr}, where we compare our numerical results to a scaling function involving a single fit parameter $D_2$,
\begin{equation}
\sqrt{D_2 t}\,f_j(t)\approx G(y)- \dfrac{1}{96\sqrt{D_2 t}}\,{} _1F_1\left[\alpha,\frac{1}{2},-\frac{y^2}{4}\right],
\label{eq:corr2_v2}
\end{equation}
following from Eq. \ref{eq:corr2} using $\lim_{\alpha\rightarrow 2}\sin(\alpha\pi)\Gamma(1-2\alpha)=\pi/12$. Furthermore, according to  Eq. \ref{eq:corr2} the approach to Gaussian scaling is algebraically slow with exponent $1.5-\alpha\rightarrow 0$ for $\alpha\rightarrow 1.5$. As we show in the following,  at this special value $\alpha=1.5$ this algebraic convergence to scaling is replaced by a persistent logarithmic correction to the scaling ansatz.
 
\textbf{Crossover point $\alpha=1.5$.--}  In the limit $\alpha\rightarrow1.5$, both prefactors in Eq.~\eqref{eq:rates}, $c_\alpha$ and $1/(3-2\alpha)$, diverge with their difference remaining finite, $-c_\alpha+1/(3-2\alpha)\rightarrow \gamma-3/2\approx -0.92$, with $\gamma$ denoting the Euler-Mascheroni constant, resulting in a  Gaussian leading order term, $\partial_t f(k)\approx -0.92\, \lambda\, k^2$. However, an additional logarithmic correction from $\lim_{\alpha\rightarrow 1.5}c_\alpha (|k|^{2\alpha-1} -k^2) = 0.5k^2\log k^2$ also contributes. Following the derivation in Ref.~\cite{zarfaty_infinite_2019}, we rescale $k$ as $k\rightarrow k\sqrt{\lambda t\,\Omega(\lambda t)/2}$, with $\Omega(\lambda t)$ a function to be determined, and get
\begin{equation*}
\sqrt{\lambda t\,\Omega(\lambda t)/2}f_{j}(t)\approx\frac{1}{4}\int \frac{dk}{2\pi}\exp(ik\tilde y)\exp\left(\frac{k^2}{ \Omega(\lambda t)}\ln\left(2 \frac{\exp(2\gamma-3)}{\lambda t\,\Omega(\lambda t)}\right)+\frac{k^2}{\Omega(\lambda t)}\ln(k^2)\right)
\end{equation*}
with scaling variable $\tilde y=j/\sqrt{\lambda t\,\Omega(\lambda t)/2}$. The function $\Omega(\lambda t)$ is chosen in such a way that  the first term in the exponent is equal to $-k^2$, reproducing the leading order Gaussian behavior~\cite{zarfaty_infinite_2019}. This leads to
\begin{equation}
\Omega(\lambda t)=\left| W_{-1}\left(-\frac{2\exp{(2\gamma-3)}}{\lambda t} \right)\right|,
\label{eq:Omega_lambert}
\end{equation}
with $W_{-1}$ the secondary branch of the Lambert W-function. As discussed in Ref.~\cite{zarfaty_infinite_2019}, this gives $\Omega(\lambda t)\approx \ln(\lambda t)\sim \ln t$, for $t\rightarrow \infty$, yielding a logarithmic correction to the scaling ansatz.  Finally, we expand the resulting expression for large time $\Omega(\lambda t)\sim\ln(\lambda t)\gg 1$ and arrive at
\begin{align}
\sqrt{\lambda t\,\Omega(\lambda t)/2} f_{j}(t)&\approx\frac{1}{4}\int \frac{dk}{2\pi} \exp(ik\tilde y)\exp(-k^2)\left(1+ \frac{k^2}{\Omega(\lambda t)}\ln(k^2)\right)\nonumber\\
&= G(\tilde y)\left(1+\frac{1}{4\,\Omega(\lambda t)}\left(\left(-2+\tilde y^2\right)\left(-2+\gamma+\ln(4)\right)+2\exp\left(\tilde y^2/4\right){}_1F_1^{(1,0,0)}\left[\frac{3}{2},\frac{1}{2},-\frac{\tilde y^2}{4}\right]\right)\right),
\label{eq:corr15}
\end{align}
 with the superscript $(1,0,0)$ denoting the derivative with respect to the first argument. This expression shows the logarithmically slow convergence towards the Gaussian scaling function for $\alpha=1.5$, as well as a persistent logarithmic correction to the scaling form, with scaling variable $\tilde{y}= j/\sqrt{\lambda t\,\Omega(\lambda t)}.$ Furthermore, for finite $t$, the above function exhibits a heavy tail $\sqrt{\lambda t\,\Omega(\lambda t)}\,f_{j,\alpha=1.5}(t)\sim y^{-3}$ and matches our 2PI results as shown in Fig.~\ref{fig:scalingcorr}, using the single fitting parameter $\lambda$. As times were note large enough in the simulations to be in the regime where $\Omega(\lambda t)\approx \ln(t)$, we used the full expression in Eq.~\ref{eq:Omega_lambert} for $\Omega(\lambda t)$ to fit the unrescaled $f_j(t)$ at a fixed time $t$.

\textbf{Superdiffusive regime, $\alpha<1.5$.--} While there are no qualitative corrections to the scaling function in this regime, the term $\sim k^2$ in Eq.~\ref{eq:rates} leads to a correction to the exponent of the hydrodynamic tail as $\alpha\nearrow 1.5$.  When evaluating the Fourier transform numerically with the full expression in Eq.~\eqref{eq:rates} for $\alpha\lesssim1.5$, we still find an approximate hydrodynamic tail with a modified exponent reproducing the finite-time dTWA and spin-2PI results more closely than the 'bare' expression $\beta_\alpha$ and hence accounting for the slight deviations between the numerical results and $\beta_\alpha$ mentioned in the main text. For example, we get $\beta_{\alpha=1.5}\approx 0.57$ from this procedure, in agreement with 2PI ($\beta_{\alpha=1.5}^{\mathrm{2PI}}\approx 0.58\pm0.02$) and dTWA ($\beta_{\alpha=1.5}^{\mathrm{dTWA}}\approx 0.59\pm0.02$).

\section{Classical master equation in dimension $d>1$}
In the following, we extend the results of the classical Master equation to spins at locations $\mathbf{r}_i$ in $d$ dimensions. The Master equation~\eqref{eq:ME} then reads
\begin{equation}
\partial_t f_{\mathbf{r}_j} = \sum_{i\neq j} W_{|\mathbf{r}_i-\mathbf{r}_j|}(f_{\mathbf{r}_i}-f_{\mathbf{r}_j}),\qquad\mathrm{with}\qquad W_{|\mathbf{r}_i-\mathbf{r}_j|}=\frac{\lambda}{|\mathbf{r}_i-\mathbf{r}_j|^{2\alpha}}.
\end{equation}
Fourier transforming again diagonalizes the differential equation, yielding
\begin{equation}
f(|\mathbf{k}|,t)=f(|\mathbf{k}|,0)\exp{\left\lbrace \left(W(|\mathbf{k}|)-W(|\mathbf{k}|=\mathbf{0})t\right)\right\rbrace}.
\end{equation} 

\textbf{Two spatial dimensions d=2.--} Denoting $k\equiv|\mathbf{k}|$ in the following, we evaluate the Fourier transform of the transition amplitudes in continuous space with both an IR (system size $L$) and UV (lattice spacing $a=1$) cutoff, yielding
\begin{align}
W(k)-W(k=0) &= \lambda \int_1^L \mathrm{d}r\,r \int_0^{2\pi} \mathrm{d}\theta \left(e^{-ikr\cos(\theta)}-1\right)\frac{1}{r^{2\alpha}}\\
&= \lambda\,2\pi \int_1^L \mathrm{d}r\, r^{1-2\alpha} \left(\mathcal{J}_0(kr)-1\right),
\end{align}
with $\mathcal{J}_0(kr)$ denoting the zeroth order Bessel function of the first kind. For $\alpha<1$ we get a divergence in the thermodynamic limit  $L\rightarrow\infty$, hence we expect the dynamics to be described by the infinite ranged mean field model in that regime. Concentrating on $\alpha\geq1$, we can remove the IR cutoff and arrive at
\begin{align}
W(k)-W(k=0) &\approx 2\pi\lambda  \int_1^\infty \mathrm{d}r\, r^{1-2\alpha}\left[ \mathcal{J}_0(kr) - 1\right]\\&\approx 2\pi\lambda \left[k^{2\alpha-2}\int_0^\infty \mathrm{d}x\, x^{1-2\alpha}\left( \mathcal{J}_0(x) - 1\right) + \dfrac{k^2}{4}\int_0^1 r^{3-2\alpha} \mathrm{d}r \right],
\end{align}
\begin{equation}
W(k)-W(k=0) \approx \lambda \left[-c_\alpha k^{2\alpha-2} +\frac{k^2\,\pi}{2(4-2\alpha)}\right],
\end{equation}
with
\begin{align}
c_\alpha&=2\pi\int_0^\infty \mathrm{d}x\, x^{1-2\alpha}\left(1- \mathcal{J}_0(x) \right)= -\frac{2^{2-2\alpha}\,\pi\Gamma(1-\alpha)}{\Gamma(\alpha)}\quad\mathrm{for}\quad1<\alpha<2.
\end{align}
We see that a superdiffusive solution is obtained for $1<\alpha\leq 2$, where the first term $\sim k^{2\alpha-2}$ dominates. Neglecting all other terms, we hence arrive at the scaling ansatz for a localized excitation $f(k,t=0)=\frac{1}{4}$
\begin{equation}
f_\mathbf{r}(t)=(\lambda c_\alpha\,t)^{-\frac{2}{2\alpha-2}}F^{2D}_\alpha\left(\frac{|\mathrm{r}|}{(\lambda c_\alpha\,t)^\frac{1}{2\alpha-2}}\right),
\end{equation}
with
\begin{equation}
F^{2D}_\alpha(y)= \frac{1}{8\pi}\int_0^\infty\mathrm{d}k\,k\, \mathcal{J}_0(ky)\, e^{- k^{2\alpha-2}}.
\end{equation}

\textbf{Three spatial dimensions d=3.--} We similarly get
\begin{align}
W(k)-W(k=0) &= 2\pi\lambda \int_1^L \mathrm{d}r\,r^2 \int_0^{\pi} \mathrm{d}\theta\,\sin{\theta}\left(e^{-ikr\cos(\theta)}-1\right)\frac{1}{r^{2\alpha}}\\
&= 4\pi\lambda\, \int_1^L \mathrm{d}r\, r^{2-2\alpha} \left(\frac{\sin{(kr)}}{kr}-1\right),
\end{align}
where we get an IR divergence and hence expect mean-field behavior for $\alpha<3/2$. Considering only $\alpha\geq 3/2$, we set $L\rightarrow\infty$ and get for small $k$
\begin{equation}
W(k)-W(k=0) \approx \lambda \left[-c_\alpha k^{2\alpha-3} +\frac{k^2\,2\pi}{3(5-2\alpha)}\right],
\end{equation}
with
\begin{equation}
c_\alpha=-4\pi\sin(\pi\alpha)\Gamma(2-2\alpha)\quad\mathrm{for}\quad1.5<\alpha<2.5.
\end{equation}
Now superdiffusive behavior is seen for $1.5<\alpha<2.5$ with a scaling ansatz in real space for a localized excitation
\begin{equation}
f_\mathbf{r}(t)=(\lambda c_\alpha\,t)^{-\frac{3}{2\alpha-3}}F^{3D}_\alpha\left(\frac{|\mathrm{r}|}{(\lambda c_\alpha\,t)^\frac{1}{2\alpha-3}}\right),
\end{equation}
with
\begin{equation}
F^{3D}_\alpha(y)= \frac{1}{8\pi^2y}\int_0^\infty\mathrm{d}k\,k \sin(ky)\, e^{- k^{2\alpha-3}}.
\end{equation}

\section{Integrable limit $\alpha\rightarrow\infty$}

The long-range XY model in Eq.~1 of the main text converges to an integrable point with increasing exponent $\alpha\rightarrow\infty$, where the diffusive hydrodynamic description is expected to break down. Here we discuss the influence of the vicinity of this integrable point on the spin transport.

The integrable point at $\alpha\rightarrow\infty$ corresponds to the nearest-neighbor interacting XY model with Hamiltonian
\begin{equation}
\hat H = -\frac{J}{2}\frac{1}{\mathcal{N_{\alpha\rightarrow\infty}}}\sum_{\langle i,j\rangle} \left(\hat{S}^x_i\hat{S}^x_j+\hat{S}^y_i\hat{S}^y_j\right).
\end{equation}
Using $\mathcal{N_{\alpha\rightarrow\infty}}=\sqrt{2}$ and applying a Jordan-Wigner transformation $\hat{S}^x_i=\frac{1}{2}(\hat c^\dagger_i+\hat c_i)$, $\lbrace\hat c_i, \hat c_i^\dagger\rbrace=1$, we arrive after a Fourier transformation $\hat{c}_j=\sum_k e^{ikj}\hat c_k/\sqrt{L}$ at
\begin{equation}
\hat H=-\frac{J}{\sqrt{2}} \sum_k \cos(k)\hat{c}_k^\dagger \hat{c}_k.
\end{equation}
This means that for $\alpha\rightarrow\infty$ we expect ballistic spin transport with a velocity given by the group velocity $v_g=\mathrm{max}_k (\partial_k (J\cos(k)/\sqrt{2}))=J/\sqrt{2}$.Note that while we employed periodic boundary conditions here, the same result could have been obtained with open boundary conditions where the eigenfunctions are not plain waves of the form $e^{ikj}$ but standing waves $\sim \sin(kj)$ with $k=n\pi/(L+1), n\in \lbrace 1,..,L\rbrace$~\cite{jurcevic_quasiparticle_2014}.

\begin{figure*}[t!]
\includegraphics{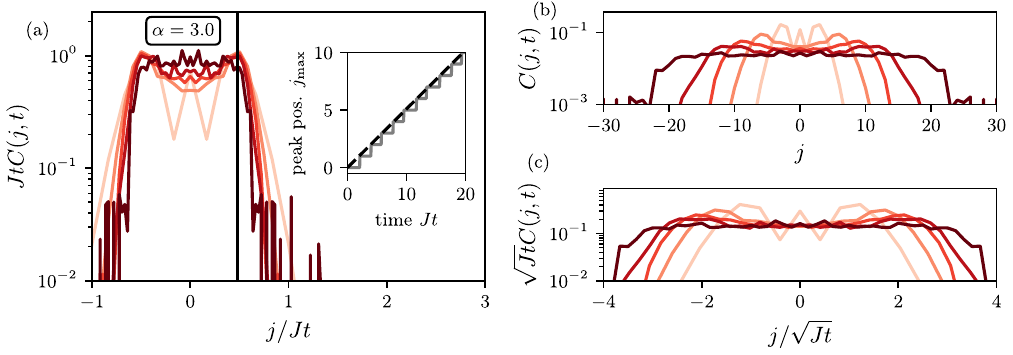}
\caption{\textbf{Ballistic spreading at short times for $\alpha=3$ from dTWA.--}  At short times ($Jt\in\lbrace 6,12,18,24,36\rbrace$) the spin correlation function shows ballistic peaks propagating towards the boundaries of the chain. (a) The scaling collapse of $C(j,t)$ when rescaling with $Jt$ as well as a linear growth of the peak position (inset) indicates ballistic transport. The black line indicates the velocity $v\approx0.5J$ extracted from the growth of the peak position (dashed line in the inset). (b) Unrescaled plot showing the propagation of the peaks. (c) Diffusive rescaling showing the late time crossover to diffusive behavior at the center of the chain.  \label{fig:integrablelim}}
\end{figure*}

In Fig.~\ref{fig:integrablelim} we show the spin correlation function $C(j,t)$ as defined in the main text obtained from dTWA for $\alpha=3$. For short times, we find ballistically propagating peaks that get gradually damped as they move towards the edges of the chain. From the time evolution of the position of the peaks we can deduce the propagation velocity and find $v_{\alpha=3}\approx 0.5J$, which is not too far from the nearest-neighbour result of $\sqrt{2}J\approx 0.7J$. As we found this discrepancy not to change as $\alpha$ is increased, we interpret it as a short-coming of this method, which is expected to work less well as the interactions become shorter ranged~\cite{rey_2015}. 

At later times, the center of the correlation function again indicates diffusive scaling, showing that interactions between particles are still strong enough to effectively dephase the system, leading to classical hydrodynamical transport at late times. We expect the time at which diffusive transport is restored to diverge as $\alpha\rightarrow\infty$. Whether the crossover to undamped ballistic transport happens at any finite $\alpha$, i.e. whether the long range interacting XY model becomes integrable at $\alpha<\infty$ is an open question.

We do not find any such peaks revealing the nearby integrable point in spin-2PI simulations, in line with the previous finding that this method is not able to capture integrable dynamics in the XXZ spin chain~\cite{PhysRevB.98.224304}.

\section{Spin conductivity}

In this section we examine the spin conductivity $\sigma(q,\omega)$, as calculated from linear response theory, and show that the DC conductivity $\sigma_{\mathrm{DC}}=\lim_{\omega\rightarrow 0}\lim_{q\rightarrow 0} \sigma(q,\omega)$ diverges in the superdiffusive region $\alpha<1.5$.

First we deduce $\sigma(q,\omega)$ in frequency space from the spin correlation function $C(q,t)$ in real time. We assume that $C(q,t)$ decays as
\begin{equation}
C(q,t)=C(q)\exp(-D_\alpha |q|^{2\alpha-1}t),
\label{eq_C_supp}
\end{equation}
supported by our results in the main text. For the initial state discussed there, $C(q)\equiv C(q,t=0)=0.25$, which coincides with the spin susceptibility at infinite temperature.

Performing a Laplace transform $\tilde C(q,z)=\int_0^\infty dt e^{izt} C(q,t)$, we arrive at
\begin{equation}
\tilde C(q,z)=\frac{i}{z+iD_\alpha |q|^{2\alpha-1}} C(q).
\end{equation}
The above equation makes the crossover from diffusive over ballistic to superballistic transport explicit as the power of $|q|$ in the pole of $\tilde C(q,z)$ determines this property. Also note that there is no damping of these hydrodynamic modes.

Moreover, one can show that $C(q,\omega)=2 \mathrm{Re}(\tilde C(q,z+i0^+)$ \cite{Forster} from $\tilde C(q,z)=\int\frac{\mathrm{d}\omega}{2\pi i} C(q,\omega)/(\omega-z)$ and the fact that $C(q,t)$ is real and even (although not explicit in Eq.~\ref{eq_C_supp}, which is only defined for $t>0$, this may be seen from $C(x,t)=\Tr(S^z(x,t)S^z(0,0))=\frac{1}{2}(\Tr(S^z(x,t)S^z(0,0))+\Tr(S^z(0,0)S^z(x,t)))$.). Hence,
\begin{equation}
C(q,\omega)=C(q) \frac{2 D_\alpha |q|^{2\alpha-1}}{\omega^2-(D_\alpha |q|^{2\alpha-1})^2}.
\end{equation}

Finally, we assume that a continuity equation of the form
\begin{equation}
\partial_t S^z(x,t) + \partial_x j(x,t)=0
\end{equation}
holds, where $j(x,t)$ is the spin current density, which is in general non-local in the case of long-range interactions~\cite{arkhincheev_nonlinear_2001}. In the limit of high temperatures $T$, the spin conductivity is given by $\sigma(q,\omega)=\frac{1}{2T}\int\mathrm{d}t \int\mathrm{d}x\, e^{i\omega t-iqx}\left\langle j(x,t)j(0,0)\right\rangle$ (here we set $k_B=1$). It follows that
\begin{align}
T\sigma(q,\omega)&=\frac{\omega^2}{2q^2}C(q,\omega)\\
&=C(q) \frac{D_\alpha \omega^2|q|^{2\alpha-3}}{\omega^2-(D_\alpha |q|^{2\alpha-1})^2}.
\end{align}
For the DC conductivity we then get
\begin{equation}\label{eq:sigDC}
T \sigma_{\mathrm{DC}}=\Bigg\{\begin{array}{lr}
\infty & \text{for } \alpha<1.5\\
D_\alpha C(q)        & \text{for } \alpha= 1.5
\end{array},
\end{equation}
showing that $\sigma_{\mathrm{DC}}$ diverges in the superdiffusive regime while it follows the Einstein relation in the diffusive regime. Note that this divergence does not stem from a divergence of the $q=0$ conductivity as $\omega\rightarrow 0$ as for a normal metal, but from the fact that the conductivity diverges for any frequency $\omega$ as $q\rightarrow 0$, i.e. $\sigma(\omega,q\rightarrow 0)=\infty$. This is a result of the non-local character of spin transport in this model.

\begin{figure*}[b!]
\includegraphics{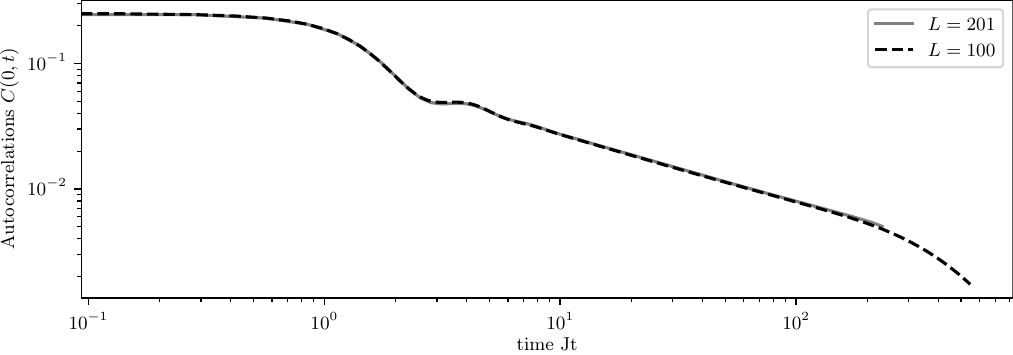}
\caption{\textbf{Comparison of hydrodynamic tail for $\alpha=2$ for biased ($L=201$) and unbiased ($L=100$) sampling from Spin-2PI.} The autocorrelation function for $L=201$ (same data as shown in Fig.~1 of the main text), where the center spin is fixed to be in spin up state agrees with the autocorrelation function for $L=100$ where all sites are sampled without bias. At late times, the algebraic decay crosses over to the finite-size exponential decay discussed in the main text for $L=100$ whereas for $L=201$ ultimately (exponentially fast) saturation to $1/(4L)$ would occur (here, the simulation had to be stopped before saturation occured). \label{fig:even sampling}}
\end{figure*}

\section{Breakdown of linear response for L\'evy flights}

In this section we argue that linear response theory breaks down for long-range interacting models displaying L\'evy flight behavior, in agreement with the discussion of the previous section. Instead, we find a non-linear relation between the (spin) current $\mathcal{J}$, and the field $E$~\cite{arkhincheev_nonlinear_2001},
\begin{equation}\label{eq:JB}
\mathcal{J}\sim E^{\,2\alpha-2}.
\end{equation}

To arrive at Eq. \eqref{eq:JB}, we add a small static homogeneous magnetic field gradient to the Hamiltonian,
\begin{equation}
\hat H(E) = -\frac{1}{2}\sum_{i\neq j=-L/2}^{L/2}\frac{J}{\mathcal{N}_{L,\alpha}|i-j|^\alpha}\left(\hat{S}^x_i\hat{S}^x_j+\hat{S}^y_i\hat{S}^y_j\right)-E\sum_{j=-L/2}^{L/2}j\,\hat{S}^z_j,
\label{eq:HamH}
\end{equation}
and we proceed by writing down a classical master equation, expected to capture the behavior of $\hat{H}(E)$,
\begin{equation}
\partial_t f_j = \sum_{i\neq j} \left[W_{i\rightarrow j}(E)\,f_i\,(1-f_j)- W_{j\rightarrow i}(E)\,f_j\,(1-f_i)\right].
\label{eq:ME_B}
\end{equation} 
As argued in the main text, according to Fermi's golden rule, $W_{i\rightarrow j}(E)$ is proportional to $|i-j|^{-2\alpha}$ based on the matrix element connecting the initial and final states. Moreover, in the presence of field $E$ and at finite temperatures $T$, the transition rates for hops to the left and right directions differ in such a way that the right hand side of Eq. \eqref{eq:ME_B} vanishes for the new equilibrium state of Hamiltonian Eq. \eqref{eq:HamH}, $f_i^{eq}(E)$. These considerations lead to a ratio determined by the different Boltzmann weights associated with these processes,
\begin{equation*}
\dfrac{W_{i\rightarrow j}(E)}{W_{j\rightarrow i}(E)}=\dfrac{f_j^{eq}(E)(1-f_i^{eq}(E))}{f_i^{eq}(E)(1-f_j^{eq}(E))}=\exp\left[(j-i)E/T\right]
\end{equation*}
resulting in a modified ansatz. In principle $\lambda\rightarrow \lambda(E)$ could still weakly depend on $(i-j)^2\,E^2$, but this would result in a subleading renormalization of the current compared to the leading order behavior discussed below.
\begin{equation*} 
W_{i\rightarrow j}(E)=\dfrac{e^{(j-i)E/(2T)}}{\cosh[(j-i)E/(2T)]}\dfrac{\lambda}{|i-j|^{2\alpha}}.
\end{equation*}
We evaluate the current response by linearizing the Fourier transform of the master equation in occupation numbers $f(k)$, resulting in
\begin{equation*}
\partial_t f(k) = [W(k;E)W(k=0;E)]\, f(k)\equiv r(k;E)\,f(k),
\end{equation*}
with a field dependent decay rate
\begin{align*}
r(k;E)&=\lambda\int_1^{L/2} dx \,\left[\dfrac{e^{xE/(2T)}}{\cosh(xE/(2T))}\dfrac{e^{-ikx} - 1}{ x^{2\alpha}}+\dfrac{e^{-xE/(2T)}}{\cosh(xE/(2T))}\dfrac{e^{ikx} - 1}{ x^{2\alpha}}\right].
\end{align*}

The broken left / right symmetry gives rise to a non-zero drift velocity, evaluated as
\begin{align}\label{eq:vdrift}
v_{\rm drift} &= i \left.\dfrac{d r(k;E)}{d k}\right|_{k=0}=2\lambda\int_1^{L/2} dx \,x^{1-2\alpha}\tanh(xE/(2T)).
\end{align}
We can distinguish three different regimes based on the behavior of $v_{\rm drift}$. For very long ranged interactions $\alpha<1$, $v_{\rm drift}$ diverges in the thermodynamic limit, resulting in a diverging current response $\mathcal{J}$ for arbitrarily small fields $E$. On the other hand, in the regime of standard diffusion, $\alpha>3/2$, Eq. \eqref{eq:vdrift} is dominated by small distances $x=O(1)$, where we can use the expansion $\tanh(xE/(2T))\approx xE/(2T)$, yielding
\begin{equation*}
v_{\rm drift}^{\rm diff}\approx \lambda E/T\int_1^{\infty}dx\,x^{2-2\alpha}=\dfrac{\lambda}{3-2\alpha}E/T=D_\alpha E/T,
\end{equation*}
with the diffusion constant defined in the main text. We thus recover the standard linear response  
$$\mathcal{J}=v_{\rm drift}\,f(k=0,0) \approx D_\alpha E /(4T),$$
yielding a DC conductivity $\sigma_{DC}\approx D_\alpha /(4T)$ in agreement with \eqref{eq:sigDC} obtained from linear response theory in the previous section.

 The two regions discussed above are separated by a regime displaying an anomalous non-linear response, $1<\alpha<3/2$. Here we can remove both the lower and upper cutoffs from Eq. \eqref{eq:vdrift}, resulting in
\begin{align*}
v_{\rm drift} &\approx 2\lambda\int_0^{\infty} dx \,x^{1-2\alpha}\tanh(xE/(2T))=2\lambda\,(E/(2T))^{2\alpha-2}\int_0^{\infty} dy \,y^{1-2\alpha}\tanh(y),
\end{align*}
indeed giving rise to anomalous scaling $\mathcal{J}\sim (E/T)^{\,2\alpha-2}$.

\section{Hydrodynamic tail for homogeneous initial states}

In the main text, we choose a Hilbert space sector with magnetization $1/2$, and also fix the centre spin to be in spin up state when sampling the trace. One may object~\footnote{We thank Achim Rosch for pointing this criticism out to us.} that this corresponds to an inhomogeneous initial state in which transport is expected to occur classically. To really show the emergence of hydrodynamics generated by fluctuations in a quantum correlation function  as done in Ref.~\cite{lux_hydrodynamic_2014}, one has to examine a homogeneous initial state. To this end here we consider the infinite temperature state studied in the main text, however sampled in an unbiased way at zero total magnetization. In Fig.~\ref{fig:even sampling} we show that the hydrodynamic tail for $\alpha=2$ shown in the main text in Spin-2PI (again, similar results are obtained in dTWA) coincides with the one obtained from an unbiased sampling of the zero magnetization sector, hence showing that the emergence of hydrodynamics discussed in the main text is not a peculiarity of the initial state chosen.
\end{widetext}
\bibliography{bib_ltt}
\end{document}